\documentstyle[psfig,aps,prb,multicol]{revtex}
\renewcommand{\narrowtext}{\begin{multicols}{2} \global\columnwidth20.5pc}
\renewcommand{\widetext}{\end{multicols} \global\columnwidth42.5pc}
\newcommand{\beq}{\begin{equation}}
\newcommand{\eeq}{\end{equation}}
\newcommand{\bea}{\begin{eqnarray}}
\newcommand{\eea}{\end{eqnarray}}

\newcommand{\calP}{{\cal P}}

\newcommand{\calV}{{\cal V}}

\newcommand{\AlGaAs}{\mbox{Al$_{x}$Ga$_{1-x}$As}}
\newcommand{\AB}{\mbox{A$_{x}$B$_{1-x}$}}

\begin{document}

\draft

\title{Optimization of semiconductor quantum devices
by evolutionary search}

\author{Guido Goldoni$^{1}$
and Fausto Rossi$^{2}$}
\address{$^1$ I.N.F.M. and Dipartimento di Fisica, Universit\`a di Modena,
Via Campi 213/A, I-41100 Modena, Italy}
\address{$^2$ I.N.F.M. and Dipartimento di Fisica, Politecnico di Torino,
Corso Duca degli Abruzzi 24, I-10129 Torino, Italy}

\date{\today}
\maketitle

\begin{abstract}

A novel simulation strategy is proposed to search for semiconductor
quantum devices which are optimized with respect to required
performances. Based on evolutionary programming, a tecnique
implementing the paradigm of genetic algorithms to more complex data
structures than strings of bits, the proposed algorithm is able to
deal with quantum devices with preset non-trivial constraints
(transition energies, geometrical requirements, etc.). Therefore, our
approach allows for automatic design, thus avoiding costly by-hand
optimizations.  We demonstrate the advantages of the proposed
algorithm by a relevant and non-trivial application, the optimization
of a second-harmonic-generation device working under resonance
conditions.

\end{abstract}


\narrowtext

In addition to the new fundamental physics which can be investigated
in lower-than-3 dimensions \cite{LD}, the enormous interest in
semiconductor nanostructures has been strongly driven by the
possibility of engineering semiconductor-based quantum devices
(SBQD's) \cite{Capasso99}; 
In particular, among the most successful applications one should mention
unipolar (i.e., intersubband) lasers \cite{Science94}---the so-called
quantum-cascade lasers--- as well as a variety of semiconductor-based
photodetectors \cite{Capasso83}.

The design and optimization of SBQD's is often grounded on an
expensive trial-and-error process: after a ``promising'' structure has
been identified, an {\em ad-hoc} optimization is performed 
with respect to a few structural
parameters. In this Letter we propose 
a novel
simulation strategy based on evolutionary programming (EP) \cite{EP},
to efficiently search in a large parameter space 
for optimal SBQD's with respect to required performances:
quantum-device optimization by evolutionary search
(Q-DOES). 

The optimization of SBQD's is a complex multidimensional maximization
problem of a fitness function containing the quality factors of the
device; the hardness of the problem grows rapidly with the number of
free parameters (number and width of the constituent layers, alloy and
dopant concentrations, etc.); furthermore, it may be required that the
device works under certain conditions (e.g., with a high optical
efficiency in a well-defined energy range) leading to a non-trivial
constrained problem. In this respect, genetic algorithms (GAs)
\cite{GAED} are often invoked for complex optimization tasks; in a
typical GA, a pool of possible solutions to the problem ---the
{\em population}--- is evolved following a genetic paradigm: solutions
belonging to the population ---coded as strings of bits, called {\em
chromosomes}--- are allowed to mate in pairs in order to cross their
genomes (genetic crossover), while random mutations (bit flips) take
place at a given rate to ensure genetic diversity. At the heart of the
algorithm is {\em selective pressure}: the fittest chromosomes are
given more chances to mate and pass their genomes to the next
generation.

For complex data structures the bit-string representation is not
usually the most natural one. In such cases, the genetic paradigm may
be implemented more efficiently by designing context-dependent genetic
operators acting on non-conventional chromosomes representing the
``natural'' data structure of the problem with improved efficency, a
strategy often referred to as {\em evolutionary programming}
 \cite{EP,GAED}.
In EP, genetic operators may have a direct physical interpretation,
and may be optimally designed to exploit the physical insight of the
problem \cite{Deaven95}. Finally, in EP inclusion of constraints is
usually simpler.

Let us consider a conventional multilayered heterostructure 
formed by a fixed number, $N_l$, of layers of a \AB\ alloy; 
layer $i$ is characterized by the A concentration, $x_i$,
and the number of \AB\ monolayers, $n_i$. We map the structure onto a
chromosome, each layer being in a one-to-one correspondence with a
{\em gene}. Hence, we design each chromosome as a vector
$(g_1,\ldots,g_{N_l})$, where $g_i$ contains all informations
concerning a given layer; in the present case $g_i=(x_i,n_i)$
\cite{generalization}. Often, structural parameters have physical
limitations \cite{type-I}; such constraints define the solution
domain. It is important (and easy) to design crossover and mutation
operators which automatically produce offsprings within the solution
domain\cite{GENOCOP}.

We next discuss how to define the fitness function. For a SBQD the
quality factors can be written as a functional of the free-carrier
wavefunctions $\psi_n$ \cite{BS}. From
these ingredients, we can define a functional
$\calV[\{\psi_n\}]$ which measures the performance of the device 
to be optimized; its evaluation may
require the calculation of energy levels, optical matrix elements, 
scattering rates,
etc., depending on the specific device requirements. Non-trivial
(non-linear) constraints, such as multiple inter-subband transition
energies, may be dealt with by means of a penalty functional which is
a measure over the parameter space, $\calP[\{\psi_n\}]
\ge 0$, going to zero when all constraints are satisfied. In our
implementation we always design $\calV$ to be minimum for the fittest
structures; more specifically, we minimize
\beq
\calV\left[\{\psi_n\}\right] \times
\exp \left\{\pm\calP\left[\{\psi_n\}\right]/
\delta p(t)\right\};
\eeq
$\delta p(t)$ is a parameter which 
sets the {\em strength} of the constraints and is, in
general, a function of the simulation time $t$; $+(-)$ is taken in
the exponential factor if $\calV$ is positive (negative) definite, at
least in the region of interest.

The proposed Q-DOES algorithm proceeds as follows. We ``evolve'' a
population of $N_p$ solutions, which is randomly initialized; at each
simulation step we randomly choose $\overline{N}_p<N_p$ solutions
which undergo, in pairs, different types of crossover operations (see
Tab.\ 1) with probabilities $p_c$, designed so that if parents belong
to the desired solution domain, so do the offsprings. Then, several
mutation operators are applied (see Tab.\ 1), with probabilities
$p_m$, to randomly chosen $g_i$'s genes of the offsprings.  Each mutation
operator has a different purpose: e.g., boundary and uniform mutations
ensure ergodicity, while non-uniform mutations perform a local search
around the solution coded by the chromosomes which reach the later
stages of the simulation.  When $\overline{N}_p$ offsprings have been
generated by crossover and mutation, we convert the alloy
concentrations and layer widths coded in their chromosomes ---{\em
genotype}--- to the corresponding confinement potentials and
space-dependent effective masses ---{\em phenotype}--- and we evaluate
the fitness function of the offsprings; in the present implementation
the electron wavefunctions $\psi_n$ and energy levels $\epsilon_n$ are
calculated within the conventional effective mass approximation (EMA)
\cite{ema}. Finally, the set of $N_p+\overline{N}_p$ chromosomes are
ranked according to their fitness-function values, and the fittest
$N_p$ chromosomes are passed to the next generation.

Let us consider a technologically
relevant and highly non-trivial optimization task: the design of a
SBQD for second harmonic generation (SHG) 
working under resonance condition. Schematically, this is
an equally spaced three-level structure, where the lowest 
inter-subband transition energies, $\epsilon_2-\epsilon_1$ = 
$\epsilon_3-\epsilon_2$, are in
resonance with a pumping radiation, $\hbar\omega$, and light emission
takes place at $\epsilon_3-\epsilon_1 = 2\hbar\omega$. 
At resonance, the emitted power is
proportional to \cite{Rosencher91} $ \mu = |\mu_{12} \mu_{23}
\mu_{31}|$, where $\mu_{ij} = \int \psi_i^*(z) z \psi_j(z) dz$ are the
intersubband dipole matrix elements corresponding to the individual
optical transitions. Due to quantum confinement, these quantities can
be greatly enhanced in a properly designed asymmetric heterostructure with
respect to bulk systems; however, to maximize $\mu$ varying the
confinement potential is a challenging task, not only because the
resonance condition should be preserved in the process, but also
because $\mu$ is unbounded, and may
become very large if resonance conditions are relaxed. Therefore, this
example is ideally suited to test the proposed simulation scheme
applied to a ``hard'' problem. This SHG optimization task has been
undertaken so far only for extremely simplified potential profiles
\cite{Rosencher91} and by means of analytical methods
\cite{Tomic97}.

In order to maximize the SHG emitted power, we set $\calV=-\mu$ and we
evolve a large population of type-I ($x\le 0.4$) \AlGaAs\ heterostructures
according to the Q-DOES algorithm described above.  Resonance
conditions with the CO$_2$ laser
radiation was enforced by setting $\calP = [\left(\epsilon_2
-\epsilon_1-\hbar\omega\right)^2+\left(\epsilon_3
-\epsilon_2-\hbar\omega\right)^2]^{1/2}$. 
The optimized structure produced by the
algorithm is shown in Fig.\ 1: it corresponds to an estimated SHG
power $\mu = 4.10\,\mbox{nm}^3$, while the intersubband transition
energies fit $\hbar\omega$ within numerical accuracy; this compares
well with (and is actually $\sim 5\%$ better than) the largest
estimated value 3.91 nm$^3$ \cite{Tomic97} obtained so far,
to our knowledge, for the same $\hbar\omega$, for an idealized
continuously graded potential \cite{nota-undershoot}.  It is important
to stress that the remarkably simple structure shown in Fig.\ 1 has
been obtained with a {\em random} initial population, i.e., {\em no
guess has been made on the shape of the potential}.

The device in Fig.\ 1 is the best performing one in a small set of
optimized structures corresponding to different local minima the
alghoritm converged to during a large set of runs, with corresponding
fitness function values falling within $\sim 5\%$ from the best
value. The rate of convergence of our algorithm in a typical run is
demonstrated in Fig.\ 2, where we show the values of the fitness
function, Eq.\ (1), and of the SHG intensity, $\mu$, of the full
population.  During the initial stages, the SHG intensity oscillates
strongly, exploring a large part of the parameter space, until the
search focuses on regions where the energy constraints are
approximately satisfied. There, the exponential factor in (1) is $\sim
1$, and the SHG intensity coincides with the fitness function (in the
figure the two vertical axes are shifted for clarity). Eventually, the
distribution of the fitness function values gets narrower, as the
population becomes degenerate, i.e., all chromosomes become similar;
at this stage, we always find the constraints to be satisfied to a
very high degree.

Although demonstrated with a simple, yet realistic problem, the
algorithm is independent on the detailed structure of the genes, and
more complex SBQD's can be dealt with using basically the same scheme;
natural candidates include quantum cascade lasers, mid- and
far-infrared photodetectors, cavity resonators, etc.  Besides, unlike
recently proposed analytical optimization methods
\cite{Tomic97}, a more accurate (microscopic) description of
electronic states may be implemented using, e.g., tight-binding or
pseudo-potential Hamiltonians. Obviously, dealing with more complex
structures and/or Hamiltonians may increase the computational cost of
the algorithm (which was not an issue in the present implementation),
which is dominated by the evaluation of the fitness function; however,
it should be noted that, in addition to the rapid convergence
demonstrated above, the evaluation of the fitness function for
different chromosomes is an intrinsically parallel task which may take
advantage of a parallel architecture. 


%
%

\widetext

 \begin{table} 
 \caption{Description of crossover (C) and mutation (M)
 mechanisms} 
 \begin{tabular}{ll} 
 C one-point & parents' $g_i$'s are
 split; offsprings $g_i$'s are generated by recombination \\
 & of the different parts \\\hline 
 C global  arithmetical & offsprings' $g_i$'s are obtained as linear
 combinations of the parents' $g_i$'s  \\\hline 
 M uniform & $g_i$ is changed randomly with uniform distribution within 
 the domain \\\hline 
 M non-uniform & $g_i$ is changed randomly with a
 probability distribution which gets narrower \\
 & as the simulation
 proceeds \\\hline 
 M boundary & $g_i$ assumes values on the domain boundaries \\\hline 
 M average & in $g_i$ and $g_j$, $x_i$ and $x_j$
 are both set to $(n_i x_i + n_j x_j)/(n_i + n_j)$ \\\hline 
 M flip & $g_i$ and $g_j$ are interchanged
\end{tabular}
\end{table}

 \begin{figure}
 \noindent
 \unitlength1mm
 \begin{picture}(100,80) 
 \put(30,0){\psfig{figure=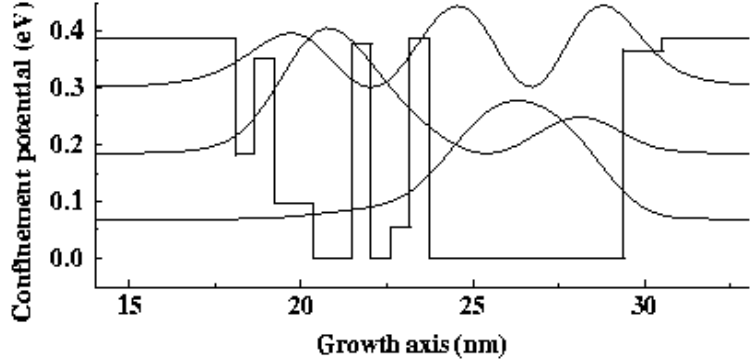,width=110mm}} 
 \end{picture}
 \vspace{6truemm}
 \caption{Potential profile and
 $|\psi_i(z)|^2$ for the lowest three confined states, shifted
 vertically by the confinement energy.  The simulation was performed with
 $N_l=30$ and $N_p=100$; Probabilities $p_c$ and $p_m$ were in the
 range [0.07,0.1]. Resonance with a pumping radiation
 $\hbar\omega=116$ meV was enforced. The estimated $\mu =
 4.10\,\mbox{nm$^3$}$ corresponds to $\mu_{12}=1.725\,\mbox{nm}$,
 $\mu_{23}=2.609\,\mbox{nm}$, $\mu_{31}=0.9104\,\mbox{nm}$.}
 \end{figure}


 \begin{figure} 
 \noindent
 \unitlength1mm
 \begin{picture}(100,110) 
 \put(30,0){\psfig{figure=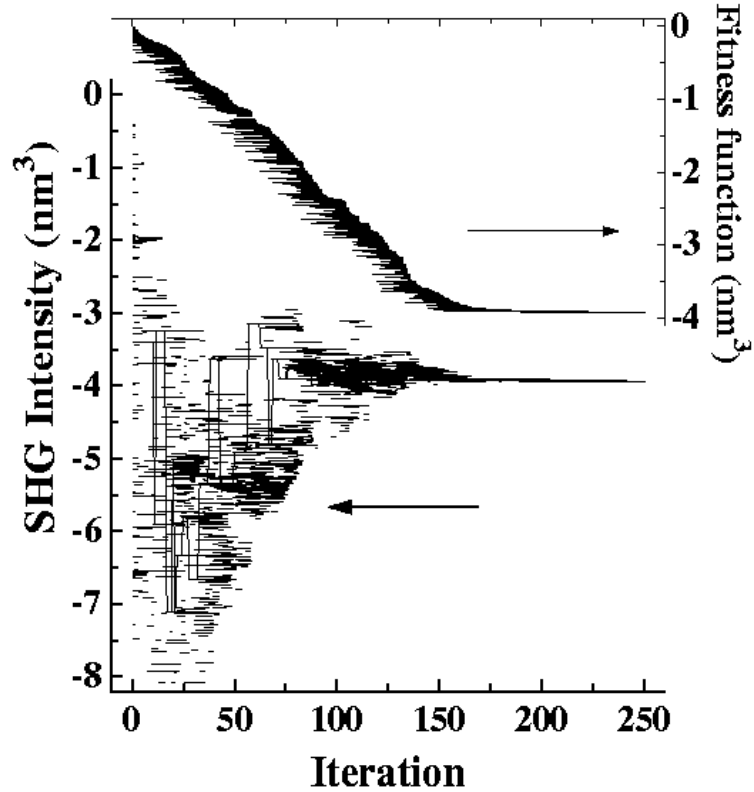,width=110mm}} 
 \vspace{6truemm}
 \end{picture}
 \caption{Fitness function values, Eq.\ (1) (top right
 axis), and SHG intensity, $\mu$ (bottom left axis), for a typical run
 (not the same leading to Fig.\ 1, but with same parameters). The two
 axis are vertically shifted for clarity. The solid line in the SHG
 intensity panel traces the fittest chromosome at each iteration.}
 \end{figure}


\begin{references}
\bibitem{generalization} 
Generalizations to more complex structures, such as ternary alloys,
doped layers, etc.\ is straightforward.
\bibitem{type-I} 
For example, for type-I \AlGaAs\ heterostructure we typically require $0<x_i<0.4$.
\bibitem{GENOCOP} In the proposed implementation, we have inspired ourselves to
the GENOCOP system described in Ref.\ \onlinecite{EP}, but additional
operators have also been introduced which are suited to the specific
problem
\bibitem{BS} 
One can write, at least in principle, any property of the quantum
structure in terms of the orthonormal basis set obtained by the
single-particle wavefunctions.
\bibitem{nota-undershoot} 
The optimized potential in Ref.\ \protect\onlinecite{Tomic97} spans
more than 0.5 eV, and cannot be implemented as a strictly type-I
structure. It is a distinct advantage of the present method the easy
implementation of physical limitations on the parameter space, such as
alloy concentrations.
\end{references}
\end{document}